\documentclass[]{pasj01}
\usepackage{bm}

\begin{document} 
\Received{}
\Accepted{}

\title{Optical polarization variations in the blazar PKS\,1749$+$096}

\author{Makoto \textsc{Uemura}\altaffilmark{1}}%
\altaffiltext{1}{Hiroshima Astrophysical Science Center, Hiroshima
  University, Kagamiama 1-3-1, Higashi-Hiroshima, 739-8526, Japan}
\email{uemuram@hiroshima-u.ac.jp}

\author{Ryosuke \textsc{Itoh}\altaffilmark{2}}
\altaffiltext{2}{Department of Physics, Tokyo Institute of Technology,
  2-12-1 Ohokayama, Meguro, Tokyo 152-8551, Japan}


\author{Ioannis \textsc{Liodakis}\altaffilmark{3}}
\altaffiltext{3}{Department of Physics and Institute for Theoretical
  and Computational Physics (ITCP), University of Crete, 71003,
  Heraklion, Greece}

\author{Dmitry \textsc{Blinov}\altaffilmark{3,4}}
\altaffiltext{4}{Astronomical Institute, St. Petersburg State
  University, Universitetsky pr. 28, Petrodvoretz, 198504
  St. Petersburg, Russia}

\author{Masanori \textsc{Nakayama}\altaffilmark{5}}
\altaffiltext{5}{Department of Information and Computer Science, Keio
  University 3-14-1 Hiyoshi, Kohoku-ku, Yokohama 223-8522, Japan}

\author{Longyin \textsc{Xu}\altaffilmark{5}}

\author{Naoko \textsc{Sawada}\altaffilmark{5}}

\author{Hsiang-Yun \textsc{Wu}\altaffilmark{5}}

\author{Issei \textsc{Fujishiro}\altaffilmark{5}}


\maketitle

\begin{abstract}
We report on the variation in the optical polarization of the blazar
PKS\,1749$+$096 observed in 2008--2015. The degree of polarization
(PD) tends to increase in short flares having a time-scale of a few
days. The object favors a polarization angle (PA) of
$40^\circ$--$50^\circ$ at the flare maxima, which is close to the
position angle of the jet ($20^\circ$--$40^\circ$). Three clear
polarization rotations were detected in the negative PA direction
associated with flares. In addition, a rapid and large decrease in the
PA was observed in the other two flares, while another two flares
showed no large PA variation. The light curve maxima of the flares
possibly tend to lag behind the PD maxima and color-index minima. The
PA became $-50^\circ$ to $-20^\circ$ in the decay phase of active
states, which is almost perpendicular to the jet position angle. We
propose a scenario to explain these observational features, where
transverse shocks propagate along curved trajectories. The favored PA
at the flare maxima suggests that the observed variations were
governed by the variations in the Doppler factor, $\delta$. Based on
this scenario, the minimum viewing angle of the source,
$\theta_\mathrm{min}=4.8^\circ$--$6.6^\circ$, and the location of the
source, $\Delta r\gtrsim 0.1\>$pc, from the central black hole were
estimated. In addition, the acceleration of electrons by the shock and
synchrotron cooling would have a time-scale similar to that of the
change in $\delta$. The combined effect of the variation in $\delta$
and acceleration/cooling of electrons is probably responsible for the
observed diversity of the polarization variations in the flares.
\end{abstract}

\section{Introduction}

Blazars are a sub-class of active galactic nuclei (AGN) that can be
observed if the jet axis is directed toward the Earth. The emission
from the jet is enhanced because of the Doppler beaming effect. The
relativistic beaming is also responsible for violent variability, which is
commonly observed in blazars (\cite{bla78bllac}). These features make
blazars excellent targets to understand the physics of jets. 

The radio--X-ray emission from blazars is dominated by the synchrotron
emission from jets, although the other components, such as the broad
line region and host galaxy can contaminate the
emission. Flat-spectrum radio quasars 
(FSRQs) are blazars that have emission lines originated from AGN in
the optical spectrum (equivalent width, EW~$\gtrsim 5\>$\AA). BL\,Lac
objects (BL\,Lacs), on the other hand, exhibit no or only weak
emission lines. BL\,Lacs are further divided into sub-classes
depending on the peak frequency of the synchrotron emission, i.e.,
low-, intermediate-, and high-peaked BL\,Lacs (LBLs, IBLs, and HBLs,
respectively; \cite{abd10sed}). 

In FSRQs and LBLs, the synchrotron emission is optically thin in the
optical and near-infrared (NIR) regime. As a result, high and variable
polarization can be observed in these wavebands. Recently, the
rotation, or swing, of the polarization position angle ($PA$) has
received attention as a promising probe for jet and magnetic field
structures (\cite{mar08bllac,abd103c279,mar10pks1510}). However, it
has been noted that interpretation of the observed variation in
polarization is not straightforward because apparent $PA$ rotations can
also be made by non-deterministic random variations of polarization
(\cite{ike11blazar,bli15rbp,lar16rot}). The observed polarization 
possibly consists of multiple components, which also complicates the
interpretation (\cite{uem10bayes,ike11blazar}). Therefore, to extract
meaningful information, the polarization data must be carefully
analyzed not only with respect to the time-series of the degree of
polarization ($PD$) and $PA$, but also with respect to the movement in
the Stokes $QU$ plane, on an object-by-object basis. In addition, the
polarization variation should be interpreted with other types of data,
such as variations in the total flux, color, and multi-wavelength
data. 

PKS\,1749$+$096 (also known as OT\,081 and 4C\,$+$09.57) is a BL\,Lac
object at $z=0.322$ (\cite{sti88pks1749}), of which the optical
polarization behavior is poorly known. According to \citet{ghi11sed},
the spectral energy distribution (SED) of the object can be explained
by a model with a magnetic field $B=1.5\>$G, bulk Lorentz factor
$\Gamma=10$, and viewing angle $\theta=3^\circ$. The SED suggests an
LBL nature, while the strong emission line (EW$=12.5\>$\AA) implies
that it may be a transition object between an FSRQ and BL\,Lac
(\cite{ghi11sed}). \citet{lur12pks1749} reported a detailed study of
very-long baseline interferometry (VLBI) observations of the
object. The position angles of the radio knots range between
$20^\circ$--$\;40^\circ$ in the downstream region of the jet, while they
exist in a wider range in the upstream region. From the motion of the
radio knots, a minimum Lorentz factor of 10.2 was estimated.
\citet{hov09flare} and \citet{lio17flare} estimated $\Gamma$ and
$\theta$ of the variable component to be
$(\Gamma,\theta)=(7.5,3.8^\circ)$ and $(7.8,2.3^\circ)$, respectively,
based on the characteristics of radio flares of the object. According
to \citet{ito16fermi}, $\gamma$-ray flares of the 
object were detected by the Large Area Telescope on the {\it Fermi
  Gamma-ray Space   Telescope} spacecraft (LAT/Fermi), while it is a
faint source at quiescence. Optical flares are associated with the
$\gamma$-ray flares. 

Early historical observations show that the optical $PD$ of
PKS\,1749$+$096 varied between a few and $\sim 10$\%, which is very
typical for blazars (\cite{kin76pol,wil80pol,imp84pol}).
\citet{bri86pol} detected a violent polarization flare from $PD\sim
10$\% to $30$\% within a period of four days. \citet{ike11blazar}
performed the first intense photo-polarimetric monitoring of this
object, and obtained 78 data points over two years. The data revealed
that a polarization rotation event was associated with a
flare. Recently, \citet{bli16rp14} and \citet{bli16rp15} also reported
optical polarization rotations of this object. These observations
suggest that PKS\,1749$+$096 is a good source to study the
polarization rotation in blazars. \citet{ike11blazar} reported that
the correlation between the total flux and the $PD$ is weak, although
they did not consider potential time-lags between these parameters. 

In this paper, we present the first detailed study of the variation in 
the optical polarization of PKS\,1749$+$096 based on the data obtained
by the Kanata 1.5 m telescope and the RoboPol polarimeter attached to
the 1.3-m telescope of Skinakas observatory
(\cite{ike11blazar,ito16fermi,kin14rbp}). The data and reduction
procedure are described in section~2 and the observational results are
reported in section~3. We discuss the implications from the results in
section~4, and summarize our findings in section~5. 

\section{Observations}

Optical and NIR photo-polarimetric observations were performed with
the 1.5-m Kanata telescope in Higashi-Hiroshima observatory and 1.3-m
telescope in Skinakas observatory.

The data obtained with Kanata are those published in
\citet{ito16fermi}. Observations and data-reduction are fully
described in \citet{ito16fermi}. Here, we give a brief overview 
of the data shown in this paper. The observations were performed with
the TRISPEC and HOWPol instruments
(\cite{wat05trispec,kaw08howpol}). Both instruments have a polarimeter
mode that uses a rotating half-wave plate and Wollaston prism. A set
of linear polarization parameters is obtained with four consecutive
exposures at half-wave plate position angles of $0.0^\circ$,
$45.0^\circ$, $22.5^\circ$, and $67.5^\circ$. The exposure time of each
frame was typically 200~s, depending on the sky conditions. $V$ and
$J$ band data were obtained simultaneously with TRISPEC, and $V$ band
data with HOWPol from 2008 to 2010. Data-reduction involved a standard
photometry procedure; after dark-subtracted and flat-fielded images
were produced, aperture photometry was performed with the {\tt APPHOT}
package in {\tt PyRAF} and differential photometry with a comparison
star taken in the same frame. Fractional Stokes parameters, $q=Q/I$
and $u=U/I$, were obtained from the photometry of ordinary and
extra-ordinary light images. $PD$ and $PA$ were calculated from $q$
and $u$: $PD=\sqrt{q^2+u^2}$ and $PA=0.5\arctan(u/q)$.  

The data in the Skinakas observatory were obtained with the RoboPol
polarimeter attached to the 1.3m telescope. The polarimeter was
specifically designed for the blazar monitoring program. It has no
moving parts besides the filter wheel in order to avoid unmeasurable
errors caused by sky changes between measurements and the non-uniform
transmission of a rotating optical element (\cite{kin14rbp}). The data
were taken in the $R$ band from 2014 and 2015. In this paper, we use
re-analyzed data which were reported in \citet{bli16rp14} and
\citet{bli16rp15}. 
 
Sixty seven sets of four variables were used in this work, i.e., $V$
band magnitude, $Q/I$, $U/I$ (or $PD$, $PA$), and the $J$ band
magnitude from MJD~54666 to 55085 obtained with TRISPEC, 5 sets of
three variables, $V$ band magnitude, $Q/I$, and $U/I$ from MJD~55274
to 55444 with HOWPol, and 46 sets of $R$ band magnitude, $Q/I$, and
$U/I$ from MJD~56775 to 57285 with RoboPol. The $V$ and $R$ bands are
so close in the wavelength domain that the difference of variability
features is not discussed in this paper.

It is difficult to identify and extract interesting patterns from such
multi-dimensional time-series data. To overcome this difficulty, a
visualization tool was developed for the blazar polarization, which is
called TimeTubes (\cite{uem16tt}). This tool enables identification of
the variations in the magnitude, color index, $Q/I$, $U/I$, and their
respective errors in one view, and facilitates noteworthy pattern
recognition. We emphasize that the most important finding in this work
was not from standard scatter plots, but from the use of TimeTubes. We
show several examples of the TimeTubes view of the data from
PKS\,1749$+$096 in Appendix 1.

\section{Results}

\begin{figure*}
\begin{center}
\FigureFile(160mm,100mm){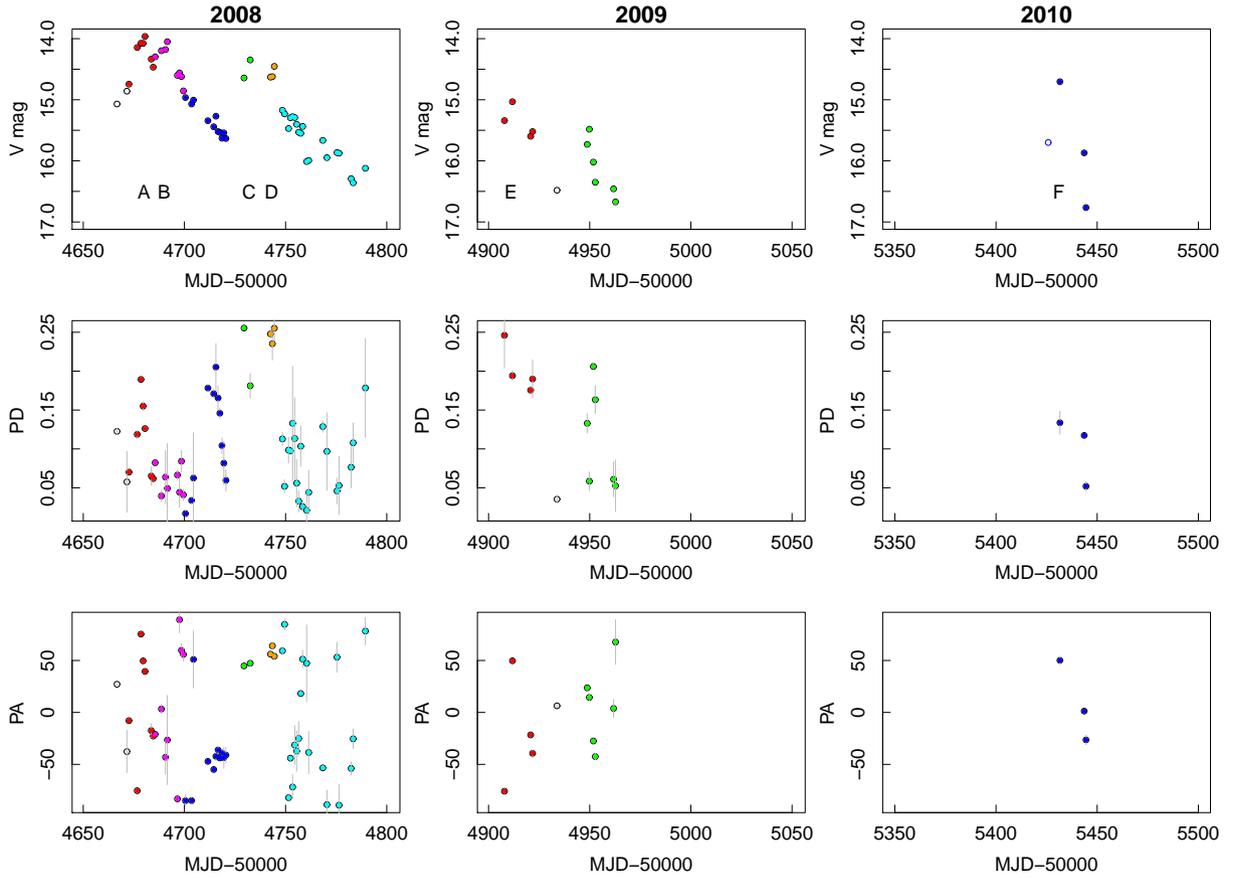}
\end{center}
\caption{$V$-band light curves (top panels), PD (middle panels), and
  PA variations (bottom panels) of PKS\,1749$+$096 observed by
  Kanata. The left, middle, and right panels are those for the data
  obtained in 2008, 2009, and 2010, respectively. The data taken
  within the same period of time is represented by the same colors in
  each year. The six flares defined in the main text are indicated by
  A to F in the top panels. The open blue symbol in the light curve in
  2010 is the converted data points in \citet{lar10atel2799} (for more
  detail, see the text).}\label{fig:lc}
\end{figure*}

\begin{figure*}
\begin{center}
\FigureFile(160mm,100mm){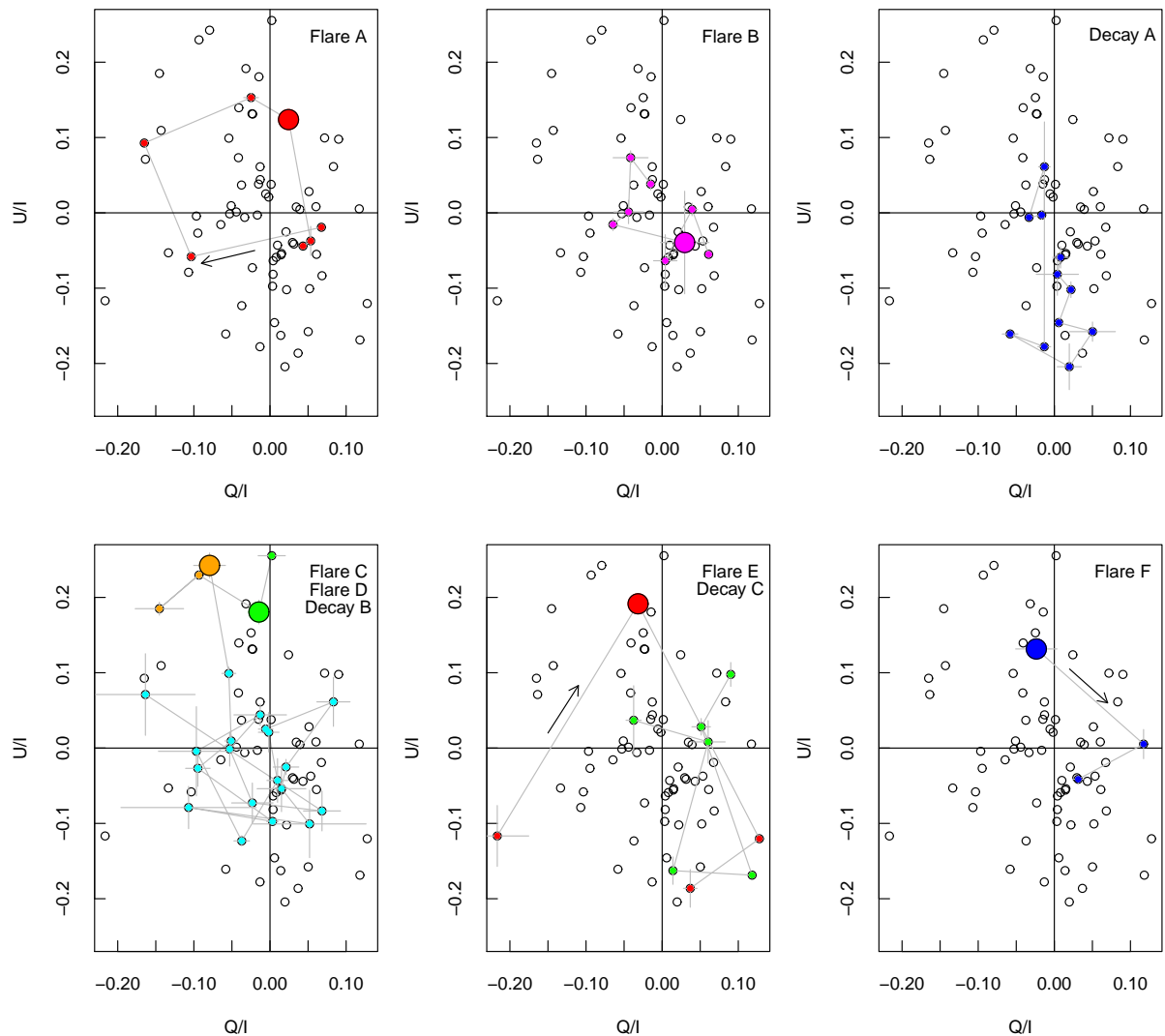}
\end{center}
\caption{The distribution and trajectories of $(Q/I, U/I)$ of
  PKS\,1749$+$096 in 2008--2010. The flares and decay phases which are
  defined in the text and indicated in figure~1 are emphasized with
  the filled circles in each panel. All data points in 2008--2010 are
  indicated by the open circles. The data taken within
  the same period of time in figure~1 is represented by the same
  colors. The large symbols indicate the flare maxima. The moving
  directions are indicated by the arrows for flare~A, E, and
  F.} \label{fig:qu}
\end{figure*}

\begin{table*}
\tbl{Features of the flares.}{%
\begin{tabular}{crrrrrr}
\hline\noalign{\vskip3pt}
Flare ID & ${T_\mathrm{max}}^*$ & ${\Delta T_{\rm c}}^\dag$ &
${\Delta T_{PD}}^\ddag$ & $PD_\mathrm{max}^\S$ &
$PA_\mathrm{peak}^\P$ & ${dPA/dt}^\Vert$\\
 & (MJD) & (day) & (day) & (\%) & (deg.) & (deg.$\>$d$^{-1}$)\\ [2pt]
\hline\noalign{\vskip3pt} 
A & 54680.52 & $3.86$ & $2.00$ & $18.9\pm 0.1$ & 
 $ 39.4\pm 0.2$ & $-16.7\pm 0.4$\\
B & 54691.55 & $6.01$ & ---     & ---           &
 --- & ---\\
C & 54732.52 & $3.06$ & $3.06$ & $25.5\pm 0.3$ &
 $ 47.3\pm 2.7$ & ---\\
D & 54744.43 & $0.00$  & $0.00$  & $25.5\pm 1.7$ &
 $ 54.1\pm 2.5$ & ---\\
E & 54911.81 & $3.95$ & $3.95$ & $24.6\pm 4.1$ &
 $ 49.7\pm 1.1$ & $-9.7\pm 0.8$\\
F & 55431.59 & ---  & $0.00$  & $13.4\pm 1.5$ & 
$ 50.1\pm 4.0$ & $-5.1\pm 1.6$\\
\hline
G & 56887.83 & ---  & $3.96$  & $24.0\pm 0.7$ &
$ 79.4\pm 1.0$ & $-10.5^{**}$ \\
H & 57213.88 & ---  & $0.00$  & $11.7\pm 0.6$ &
$ 46.1\pm 1.5$ & $-9.0^{\dag\dag}$ \\ [2pt]
\hline\noalign{\vskip3pt} 
\end{tabular}}\label{tab:flr}
\begin{tabnote}
$^*$~Times of the observed maxima in the total flux. $^\dag$~Time
  differences of the observed minima in the $V-J$ color index from
  $T_{\rm max}$. The color is not available for flare~F. $^\ddag$~Time
  differences of the observed maxima in the $PD$ from $T_{\rm
    max}$. $^\S$~Observed maximum values of the $PD$. $^\P$~$PA$ at
  $T_{\rm max}$. These three features are not given in flare~B,
  because no clear polarization flare was associated with
  it. $^\Vert$~Temporal gradient of the $PA$ for flares in which large
  variations of $PA$ were
  detected. $^{**}$\citet{bli16rp14}. $^{\dag\dag}$\citet{bli16rp15}.
\end{tabnote}
\end{table*}

Figure~\ref{fig:lc} shows the light curves, PD and PA variations of
the object observed by Kanata in 2008 (left), 2009 (middle), and 2010
(right). The data taken within the same period of time are indicated
by the same colors in each year. Table~\ref{tab:flr} lists the
features of the flares described below.

The object experienced historically bright states in 2008. It was also
in the brightest state in gamma-rays (\cite{ito16fermi}). The light
curve in 2008 can be described with the two active states around
MJD~54660--54720 and 54720--54790. In addition, short flares are
superimposed on those active states. Four short flares were
identified, flares~A, B, C, and D, as indicated in the light curve. 
The observed peak times of the $V$-band magnitude for each flare,
$T_{\rm max}$, are listed in Table~\ref{tab:flr}. 

Figure~\ref{fig:qu} shows the distribution and trajectories of
$(Q/I,U/I)$ in 2008--2010. The upper-left panel shows the trajectory
during flare~A, indicated by the red points and gray lines. The
trajectory indicates an apparent clockwise rotation in the $PA$. The
large red symbol is the data at $T_\mathrm{max}$ of the flare. The
data for flares~B, C, and D are also shown with each flare maximum
emphasized by large symbols in figure~\ref{fig:qu}. As can be seen
from those large symbols, the maxima of flares~A, C, and D have
similar $PA$s around $40^\circ$--$\;50^\circ$. The $PA$s at
$T_\mathrm{max}$, called as $PA_{\rm peak}$, are listed in
Table~\ref{tab:flr}. In contrast to flare~A, no hint of polarization
rotation was seen in our available data of flares~C and D, while those
flares were not well observed. In the left panels of
Figure~\ref{fig:lc}, the blue and cyan symbols correspond to the data 
in the fading phase from the first and second active states. We call
them decays A and B. Figure~\ref{fig:qu} shows that the polarization
of these decay phases favors negative $U/I$.  

The middle panels in Figure~\ref{fig:lc} show the data from 2009. 
Flare~E was identified, as indicated by the red symbols. The observed
maximum of this flare has a PA close to that of flare~C, as shown in
Figure~\ref{fig:qu} and Table~\ref{tab:flr}. Figure~\ref{fig:qu} shows
that the PA changed dramatically during the flare. The direction of
variation in $PA$ is again negative, as in the case of flare~A, while
the number of data is insufficient to clearly define the polarization
variation. After flare~E, the object retained a faint state with a 
possible minor flare, as shown by the green symbols in the
figures. Large negative $U/I$ values were recorded during this phase,
as observed in decay~A. 

The data from 2010 are shown in the right panels of
Figure~\ref{fig:lc}. \citet{lar10atel2799} reported an optical flare 
of this object in this year; $R=15.28\>$mag on 17 Aug. 2010. Following
this report, we began observations on 22 Aug. The object was the
brightest on this night ($V=14.70$ at 22.59 UT Aug), and then faded.
Figure~\ref{fig:lc} shows the light curve of this flare. Two more
measurements were obtained after the peak. While the $V-R$ color is
unknown during this flare, the object was presumably brighter on 22
Aug. than that on 17 Aug. because LBL have a typical color of 
$V-R\sim 0.5$, which indicates $V\sim 15.7$ on 17 Aug.
(\cite{gau12blcol}). This converted magnitude is shown as the open
blue circle in the light curve, which is called flare~F here. The $PA$
of the observed maximum of flare~F is close to that of flare~E. The
object experienced a rapid and large decrease in the $PA$ during the
fading phase of this flare, as observed in flares~A and E.

\begin{figure*}
\begin{center}
\FigureFile(107mm,107mm){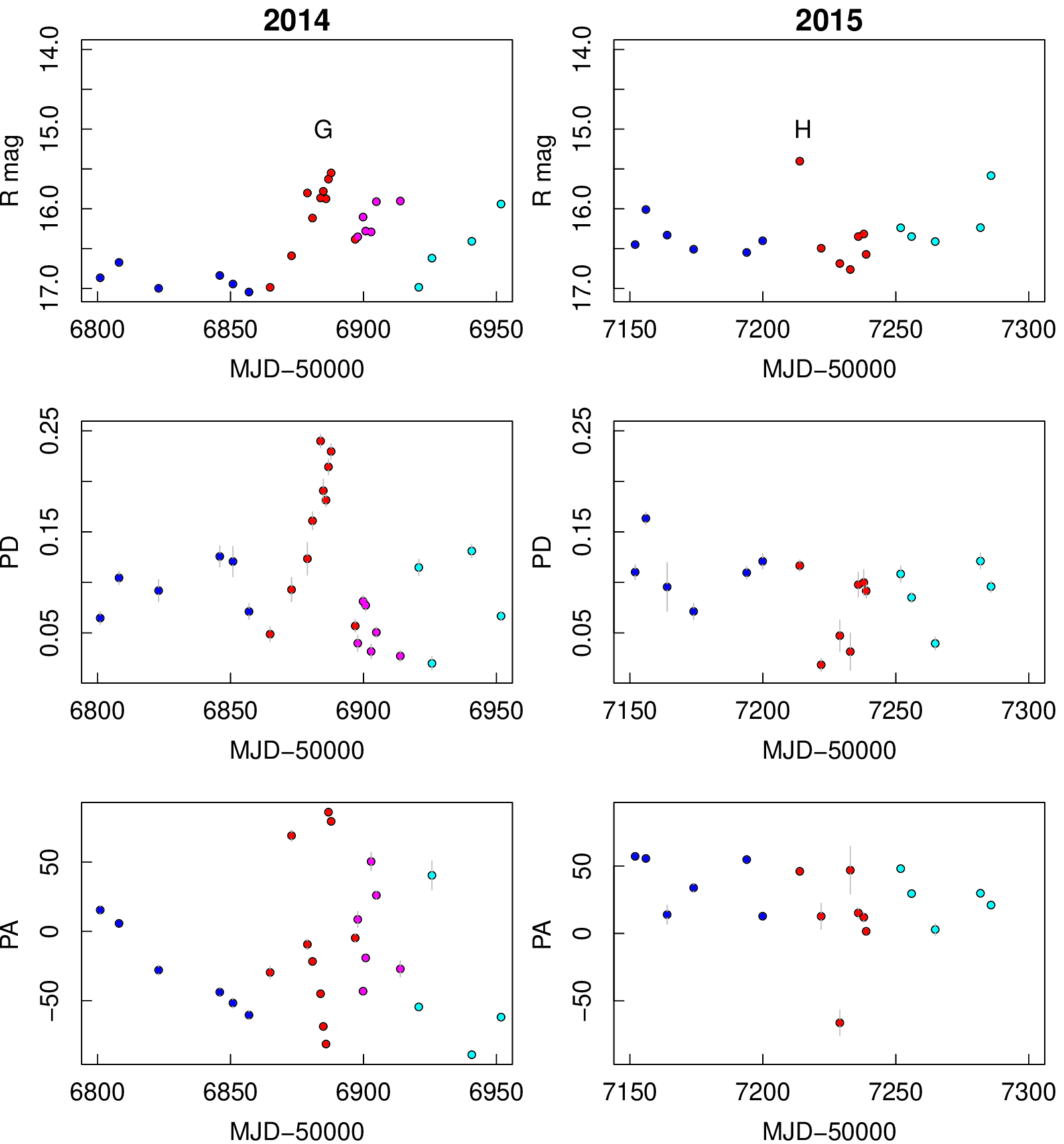}
\end{center}
\caption{$R$-band light curves (top panels), PD (middle panels), and
  PA variations (bottom panels) of PKS\,1749$+$096 observed by
  RoboPol. The left and right panels are those for the data obtained
  in 2014 and 2015, respectively. The scales and symbols are the same
  as those in Figure~\ref{fig:lc}.}\label{fig:rbplc}
\end{figure*}

\begin{figure*}
\begin{center}
\FigureFile(160mm,100mm){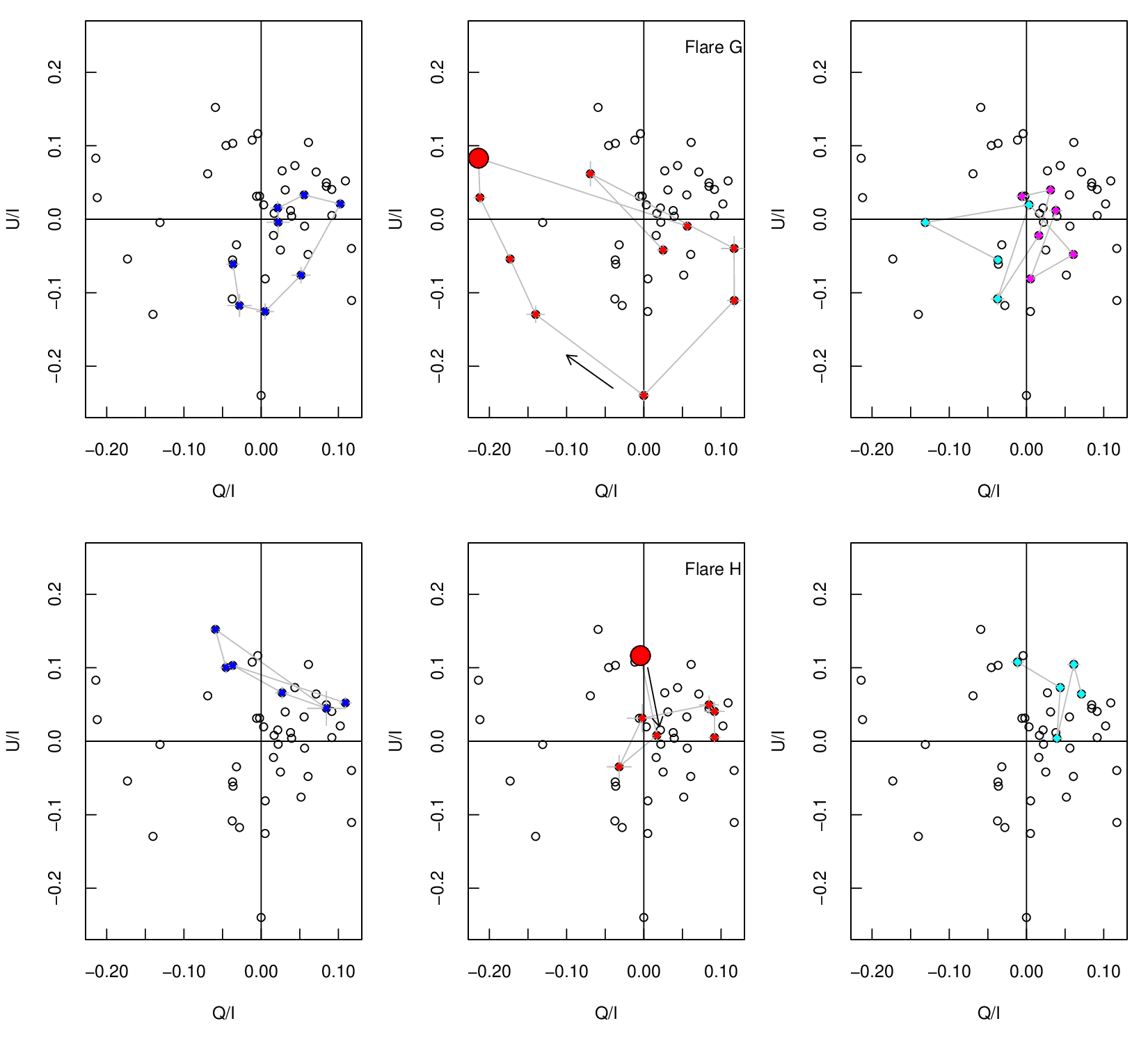}
\end{center}
\caption{The distribution and trajectories of $(Q/I, U/I)$ of
  PKS\,1749$+$096 in 2014 and 2015. The scales and symbols are the
  same as those in Figure~\ref{fig:qu}.}\label{fig:rbpqu} 
\end{figure*}

Figures~\ref{fig:rbplc} and \ref{fig:rbpqu} are the same as
Figures~\ref{fig:lc} and \ref{fig:qu}, but for the data observed by
RoboPol in 2014 and 2015. \citet{bli16rp14} and \citet{bli16rp15}
reported two polarization rotation events in this period of time. 

The first event was observed in MJD~56860--56900 with a rotation rate
of $dPA/dt=-10.5\>\mathrm{deg\>d^{-1}}$, and a $PA$ amplitude of
$335.1^\circ$ (\cite{bli16rp14}). The period of this event is
indicated by the red points in the left panels of
Figure~\ref{fig:rbplc} and upper panels of Figure~\ref{fig:rbpqu}. The
event was associated with an optical flare, of which the maximum
occurred at when $PA=79.4\pm 1.0^\circ$. We call this flare as
flare~G. The characteristics of the trajectory on the $(Q/I,U/I)$
plane are analogous to those of flares~A, E, and F, that is, a
clockwise rotation with high $PD$, as shown in
figure~\ref{fig:rbpqu}. Besides the flare, the object favors negative
$U/I$. The object was in the faintest state during our observations
before the flare. 

The second event was observed in MJD~57210--57240 with a rotation rate
of $dPA/dt=-9.0\;\mathrm{deg\>d^{-1}}$, and a $PA$ amplitude of
$224.5^\circ$ (\cite{bli16rp15}). The period is indicated by the red
points in the right panels of Figure~\ref{fig:rbplc} and lower panels
of Figure~\ref{fig:rbpqu}. This event was unique in terms of both the
light curve and polarization variation. The object kept a level
slightly brighter than the quiescence, and favors an area of positive
$Q/I$ and $U/I$ throughout 2015. These features are probably due to
the emergence of a new emitting component having the polarization of
positive $Q/I$ and $U/I$. A possible short flare, which we call
flare~H, was observed at the onset of the polarization rotation when
$PA=46.1\pm 1.5 ^\circ$. $PD$s were small during the polarization
rotation, except for the short flare. 

\begin{figure}
\begin{center}
\FigureFile(80mm,160mm){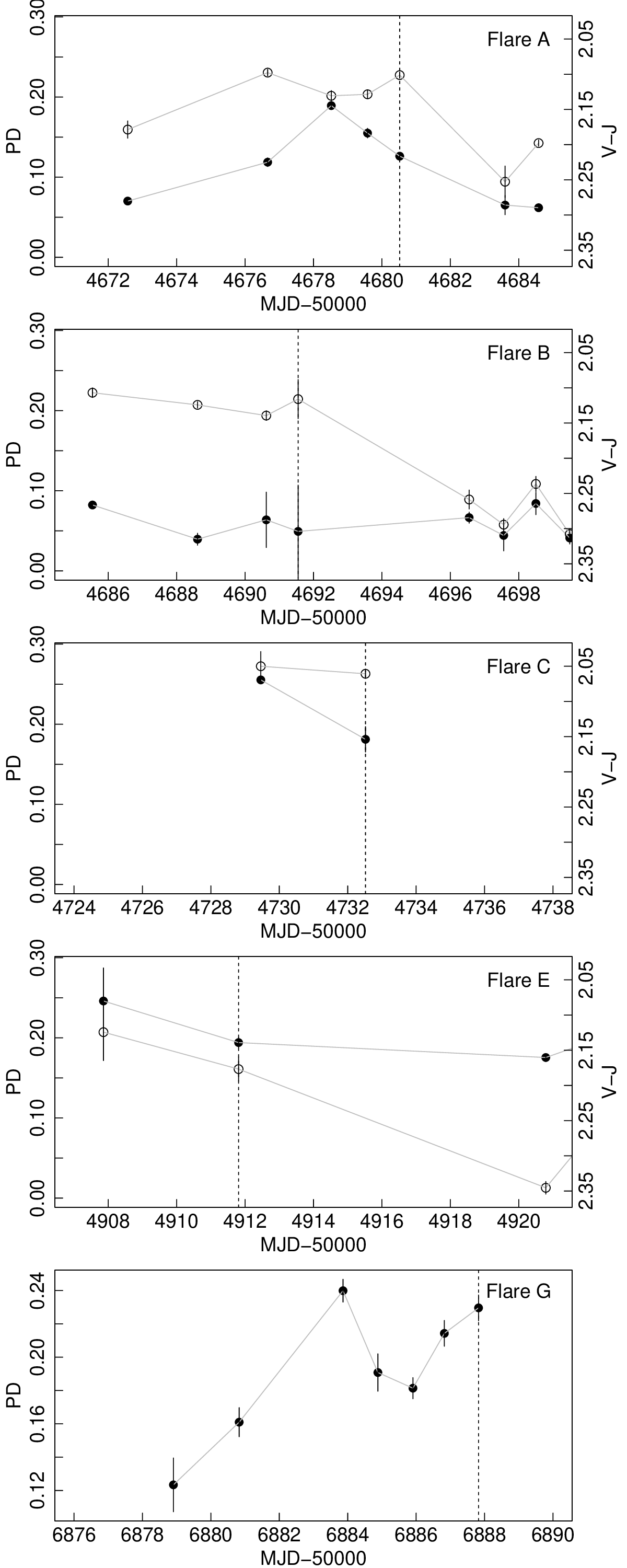}
\end{center}
\caption{PD and $V-J$ color variations of flares A, B, C, E, and
  G. The filled and open circles indicate PD and $V-J$,
  respectively. The observed maxima of each flare in the total flux
  are indicated by the dashed vertical lines. The $V-J$ data is not
  available for flare G.}\label{fig:colpd}
\end{figure}

Figure~\ref{fig:colpd} shows the $PD$ (filled circles) and $V-J$ (open 
circles) variations of flares A, B, C, E, and G. The vertical dashed
lines indicate $T_\mathrm{max}$ of each flare. $PD$ flares were
associated with all flares, except for flare~B, in which the $PD$ 
remained low throughout the flare. Flare~A was well observed in both
the rising and decaying phases, which suggests a clear time-lag of
$T_\mathrm{max}$ against the $PD$ maximum. Similarly, the observed PD
maximum precedes $T_{\rm max}$ in flares~C and E. In flare~G, the
observed maximum of $PD$ precedes $T_{\rm max}$ by $4\,$d. However,
this $PD$ peak is possibly not associated with the flare maximum
because $PD$ again increased toward $T_{\rm max}$. The time lags of
$T_\mathrm{max}$ against the observed peaks of $V-J$
($\Delta T_{\rm c}$) and $PD$ ($\Delta T_{PD}$) are listed in
Table~\ref{tab:flr}. It should be noted that the number of data except 
for flare~A is insufficient to make firm conclusions about the general
trend of the time-lag. The lacks of time-lags in flares D, F, and H
are mainly due to poorly-covered observations. In addition, those
observed time-lags listed in Table~\ref{tab:flr} possibly have large
uncertainties because they are obtained from observations with a
typical cadence of a few days and some flares were poorly observed. 

\begin{figure}
\begin{center}
\FigureFile(80mm,160mm){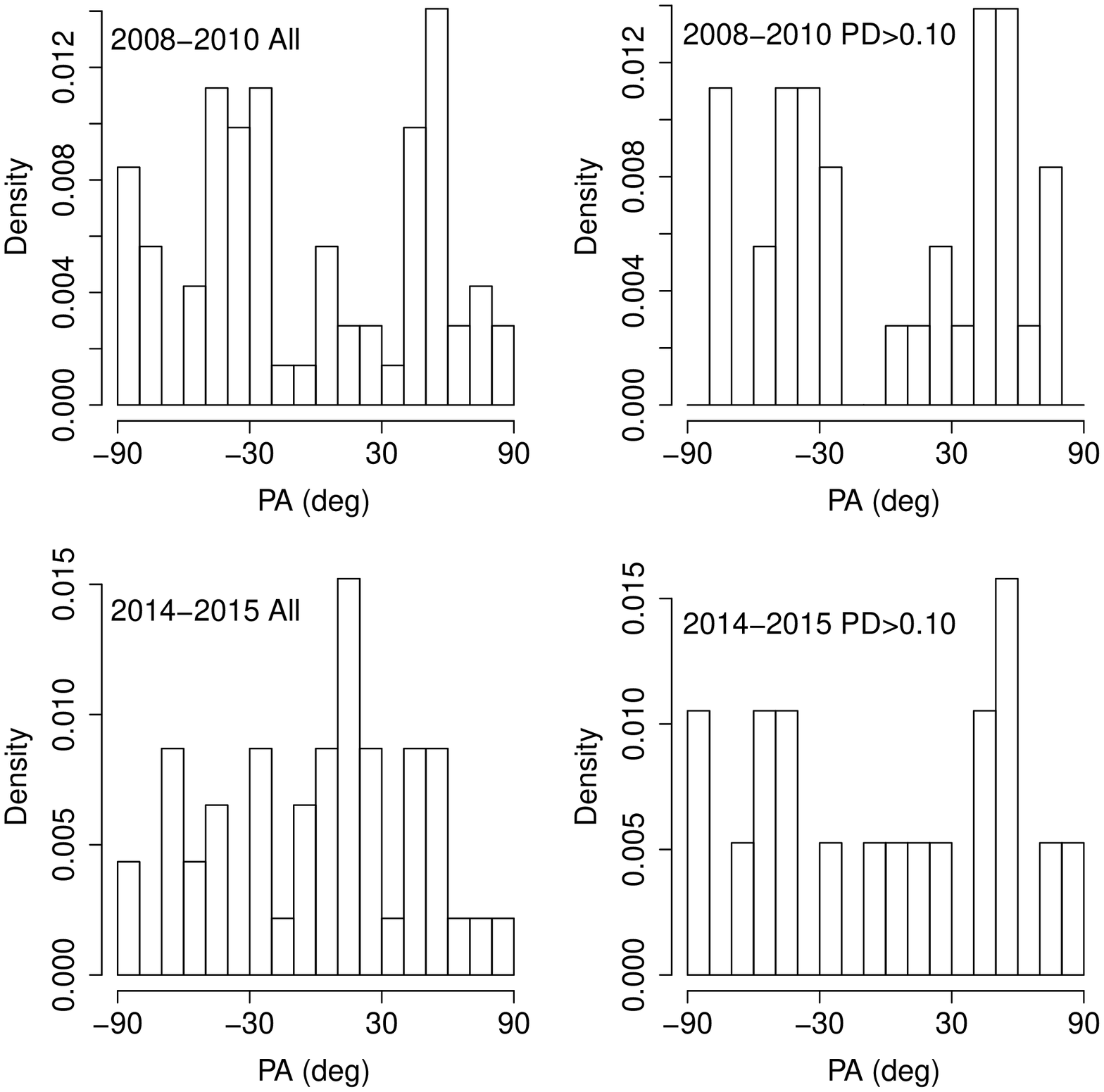}
\end{center}
\caption{Histograms of the PA for all data between 2008 and 2010
  (upper-left panel), those for PD$>0.10$ (upper-right panel), all
  data of 2014 and 2015 (lower-left), and those for PD$>0.10$
  (lower-right panel).}\label{fig:hst}
\end{figure}

Figure~\ref{fig:hst} shows histograms of the $PA$. The histogram for
all data in 2008--2010 (upper-left panel) suggests a concentration
between $PA\sim 40^\circ$--$\;50^\circ$ The other possible concentration
can be observed around $PA\sim -50^\circ$ to $-20^\circ$. The
upper-right panel of Figure~\ref{fig:hst} shows the $PA$ distribution
for the 2008--2010 data with $PD>0.10$. The non-uniformity of the $PA$
distribution is emphasized for the data with a high $PD$. The
concentration of $PA\sim 40^\circ$--$\;50^\circ$ originates from the
data around the flare maxima. The other concentration of $PA\sim
-50^\circ$ to $-20^\circ$ is due to the fading phases from the active
states. It is noteworthy that the difference between those two favored
$PA$s is $\sim 90^\circ$.

The lower panels of Figure~\ref{fig:hst} are the data from 2014 and
2015. The distribution of the all $PA$ data (left) has a possible
spike feature around $PA\sim 20^\circ$--$\;30^\circ$. This is due to the
data from 2015 which concentrates in the area of $Q/I>0$ and
$U/I>0$. The $PA$ distribution of $PD>0.10$ (right) exhibits features
similar to that in 2008--2010, that is, a concentration of
$PA\sim 40^\circ$--$\;50^\circ$. 

\section{Discussion}
\subsection{Transverse shock scenario for the short flares}

The polarization variations and rotations associated with the short
flares have common features, as shown in the previous section. First,
their values of $dPA/dt$ are of the same order of magnitude as those
shown in Table~1. Second, all detected rotations have negative
$dPA/dt$. If the rotations are made by random variations, then the
probability that five rotations are always in the negative direction
is low at $0.5^5 \sim 0.03$. Finally, among 8 flares, six flares (A,
C, D, E, F, and H) exhibit similar $PA_{\rm peak}$. The
$PA_{\rm peak}$ of Flare~G ($\sim 79^\circ$) significantly deviates
from those of the other flares. However, as noted by
\citet{bli16rp14}, the object was not observed for $9.1\>$d after the
observed maximum. During this period, the $PA$ decreased from
$79.4^\circ$ to $-4.8^\circ$. Therefore, the flare maximum could have
a similar $PA$ to those of our observations ($40^\circ$--$50^\circ$)
if the real maximum was between this  period. These common features
suggest that the polarization rotation events and flares have a common
mechanism. We discuss it in this subsection. 

We first focus on the fact that the object favors a narrow range of
$PA_{\rm peak}$. This feature suggests that the flares are mainly
caused by a geometrical effect. Here, we consider flaring sources that 
propagate along curved trajectories. In the case of the compact
emission source, the observed flux $F(t)$, can be expressed as: 
\begin{eqnarray}
  F(t)=F_0\nu^{-\alpha}\delta^{(3+\alpha)},
\end{eqnarray}
where $F_0$ and $\alpha$ are the flux in the co-moving frame of the
jet and the spectral index, respectively (\cite{der09book}). $\delta$
is the Doppler factor, which changes with time because the angle
between the velocity vector of the source and the line-of-sight is a
function of time. The flare maxima are observed at $T_\mathrm{max}$
when the viewing angle reaches the minimum at $\theta_{\rm min}$ and
$\delta$ is at maximum. A clear polarization swing (flare~A) and two
large $PA$ variations around $T_\mathrm{max}$ (flares~E and F) were
detected. These $PA$ variations are also expected in this scenario
because the direction of the magnetic field can change with the
propagation of the source along with the curved trajectory
(\cite{bjo82swing,kon85swing,nal12swing,lyu17swing}).  

\begin{figure}
\begin{center}
\FigureFile(80mm,80mm){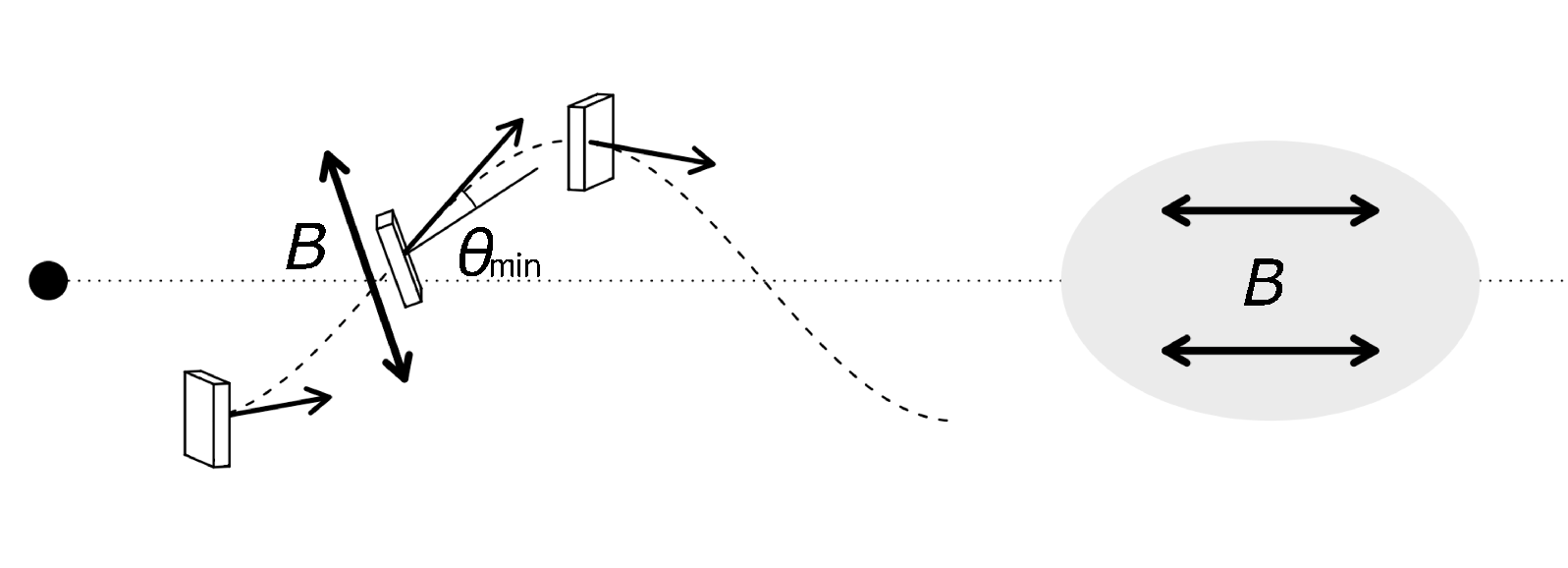}
\end{center}
\caption{Schematic view of the emitting region and magnetic field
  in the jet.}\label{fig:hel}
\end{figure}

The VLBI observations reported in \citet{lur12pks1749} show that the
position angle of the jet is $20^\circ$--$\;40^\circ$ in the downstream
region, approximately $10\>$pc from the core, while the $PA$ of the
radio knots takes a wide range of values ($-20^\circ$--$\;50^\circ$) in the
upstream region. Our observation shows that the $PA$ of
$T_\mathrm{max}$ is concentrated between $40$--$\;50^\circ$, which is
close to the jet position angle. This implies that the magnetic field
in the flaring source is almost perpendicular to the jet direction at
$T_\mathrm{max}$. Such a condition can be explained with the ordered
magnetic field in a plasma compressed by a transverse shock
(\cite{mar853c273,hug85pol}). \citet{hag08ao0235} proposed a similar
scenario for the optical flare of the blazar AO\,0235$+$164 to explain
its $PA$ close to the jet direction. 

We note that all $dPA/dt$ detected in flares~A, E, F, G, and H are
negative. This implies that the curved trajectory is governed by a
fixed structure in the jet, and is time independent, at least on a
time-scale of years. The helical magnetic field is a candidate
for such a structure, while the bending of the entire jet is also
possible if the bending structure is time independent. \citet{gab08cp}
estimated the helicity of the magnetic field in AGN jets based on the
observed rotation measure gradient and sign of parsec-scale circular
polarization. They reported that PKS~1749$+$096 exhibits a
right-handed helical magnetic field with an inward poloidal component
which corresponds to the south magnetic poles. The helicity is
consistent with the sign of $dPA/dt$ that we observed if the emitting
regions propagate along with the helical field to the downstream region
of the jet.

Figure~\ref{fig:hel} illustrates a schematic view of this
scenario. This is supported by the possible time-lag between the PD
maxima and $T_\mathrm{max}$, as shown in Table~1 and
Figure~\ref{fig:colpd}. The polarization variation of compressed
plasma in a transverse shock is predicted by several theoretical models
(\cite{bjo82swing,kon85swing,zha16shock}). According to such models,
the timing of the PD maximum can be different from
$T_\mathrm{max}$. This is because the total flux reaches the maximum
when the shocked plane is faced and the emission is only weakly
polarized. On the other hand, the PD reaches the maximum when the
shocked region is observed with a larger viewing angle and the
polarization is maximized. These polarization behavior can also be
expected in the models that only consider the geometrical effect
without compressed plasma (\cite{nal12swing,lyu17swing}). 

\citet{nal12swing} discusses the polarization variation of emitting
blobs propagating on curved trajectories. In the model, the emitting
blobs pass through the trajectories with a constant curvature radius
$R$. The flare maximum is observed when the viewing angle becomes
minimum at $\theta_\mathrm{min}$. Then, the maximum value of $dPA/dt$
and $\Delta T_{PD}$ are provided as function of $R$ and
$\theta_\mathrm{min}$, as follows:  
\begin{eqnarray}
  \frac{dPA}{dt}_\mathrm{max} &=&
  \frac{\Omega}{(1-\beta_\mathrm{blob}\cos \theta_\mathrm{min}) \sin
    \theta_\mathrm{min}} ,\\
  \Delta T_{PD} &=& \frac{1}{\Omega} \left[\mathrm{arccos}
    \left(\frac{\beta_\mathrm{jet}}{\cos \theta_\mathrm{min}} \right)
    -\beta_\mathrm{blob}\sqrt{\cos^2 \theta_\mathrm{min}
      -\beta^2_\mathrm{jet}} \right],
\end{eqnarray}
where $\Omega=\beta_\mathrm{blob}c/R$ is the angular velocity of the
blob, and $\beta_\mathrm{blob}$ and $\beta_\mathrm{jet}$ are the ratios
of the blob and jet velocities to the speed of light. As in
\citet{nal12swing}, we assume the blob and jet speeds are the same:
$\beta=\beta_\mathrm{blob}=\beta_\mathrm{jet}$. $\beta$ is given by the
Lorentz factor $\Gamma$, of the blob and jet:
$\beta=v/c=\sqrt{1-\Gamma^{-2}}$. This model is applied to our
observations of flare~A, which gives
$dPA/dt=-16.7\pm0.4\>{\rm deg}\>{\rm d}^{-1}$ and
$\Delta T_{PD}=2.00\pm1.00\>$d. Here, the uncertainty of $\Delta
T_{PD}$ is roughly estimated to be $1.00\>$d based on the observation
interval. \citet{lur12pks1749} reports the minimum Lorentz factor of
this object to be $\Gamma=10.2$ from the VLBI observations. Based on
the characteristics of radio flares, \citet{hov09flare} and
\citet{lio17flare} reports $\Gamma=7.5$ and $7.8$, respectively. For
$\Gamma=7.5$--$10.2$, the data of flare~A and Equations~(2) and (3)
provide $R=1.6$--$4.1\>$pc and
$\theta_\mathrm{min}=4.8^\circ$--$6.6^\circ$. In \citet{nal12swing},
the distance covered by the blob between the PD maxima and $T_{\rm max}$
is given as a function of $R$ and $\theta_\mathrm{min}$: 
$\Delta r_\mathrm{blob} = R \times
\mathrm{arccos}(\beta_\mathrm{jet}/\cos\theta_\mathrm{min}) = 0.1$--$0.2\>\mathrm{pc}$.

\citet{sav10beam} reported that the viewing angle $\theta$, of blazars
that are not detected in $\gamma$-rays by LAT/Fermi ranges from
$0^\circ$ to $10^\circ$ with a few exceptions having larger $\theta$,
and have a mean value of $4.4^\circ$. As mentioned in section~1,
PKS~1749$+$096 is a faint $\gamma$-ray source at quiescence. According
to \citet{ito16fermi}, it is just around the 3-sigma detection limit
in seven-day bins. The estimated $\theta_\mathrm{min}$ for flare~A 
is a typical one for non-LAT-detected blazars, and close to that
estimated in \citet{sav10beam} for PKS~1749$+$096 ($4.2^\circ$). The
estimated $\Delta r_\mathrm{blob}$ gives a lower limit of the distance
from the central black hole. The location of the flaring source at
$\gtrsim 0.1\>\mathrm{pc}$ is consistent with the standard picture of
blazars, in which the optical emitting sources are located in a
sub-parsec region. Thus, the proposed scenario can explain the
time-lag between the flare and rotation within the current
understanding of blazars. 

A problem is that the theoretical models generally provide two PD
maxima both before and after $T_\mathrm{max}$. This is because they
assume that the observed variation is only caused by the variation in
$\delta$. As a result, the PD variation is symmetric with respect to
$T_\mathrm{max}$ when symmetric trajectories on the position of
$\theta_{\rm min}$ are considered. The PD peaks were detected before
$T_\mathrm{max}$ in three flares (flare A, C, E, and possibly G, as
shown in Table~1), while no sign of the PD peak after $T_\mathrm{max}$
was observed in 6 flares. It is possible that the second PD peaks were
overlooked. Flares~A, E, and G were not observed for a few days after
$T_\mathrm{max}$, when the other PD peak may have been present.
However, this idea is not favored by the data, because there is no
observation bias to detect the first PD maximum more frequently than
the second PD maximum.

Alternatively, the lack of the second PD maximum may be explained if
the flaring sources rapidly decreased their flux density just after
$T_\mathrm{max}$. That is, we consider the temporal variation in
$F_0$ in Equation~(1). \citet{ghi11sed} reported the SED of
PKS\,1749$+$096, which suggests that the optical waveband corresponds
to the high frequency edge of the synchrotron emission. The optical
emission thus originates from the electrons with the maximum
energy. Therefore, it is possible that the decay of the flare is
governed by the synchrotron cooling. 

The synchrotron cooling time-scale of an electron in a homogeneous
magnetic field in the observer's frame $t_c$, can be estimated as
follows:
\begin{eqnarray}
  t_c = (\delta)^{-1} 5\times 10^{11} (1+z)^{1/2} B^{-3/2}\>[{\rm G}] 
  (\nu_\mathrm{obs}\>[{\rm Hz}]/\delta)^{-1/2}\> [\mathrm{s}],
\end{eqnarray}
where $B$ and $\nu_\mathrm{obs}$ are the magnetic field and
observation frequency (\cite{tuc75rad}). Based on 
the SED analysis, \citet{ghi11sed} reported $B=1.5\>$G for
PKS\,1749$+$096. \citet{lur12pks1749} estimated $\delta=10.2$--$20.4$
from VLBI observations. Using these values, we estimated
$t_c=0.03$--$0.05\>$d. The estimation of $t_c$ is highly dependent on
$B$, as evident from Equation~(4). If we use $B=0.15\>$G, which is one
order of magnitude lower than that proposed, but still acceptable for
blazars (\cite{ghi11sed}), $t_c$ is estimated to be
1.08--1.53$\>$d. These estimates suggest that the electrons
accelerated by the shock possibly lose the energy by synchrotron
cooling in a time-scale less than days. Therefore, the second PD
maximum may be missed if significant cooling starts between the first
PD maximum and $T_\mathrm{max}$. 

The observed color variations also imply that we observed the
acceleration and cooling processes by the shock during the flares. The
color of the object was bluest 3--6$\>$d before $T_\mathrm{max}$ for
flares~A, B, C, and E, as shown in Table~1. These color variations are
sometimes referred to as spectral hysteresis, i.e., a loop track in a
spectral hardness--intensity diagram
(\cite{tak96mrk421,kat00pks2155,ike11blazar}). This is explained by
the scenario that high energy electrons decay by rapid cooling, which
results in growth of the flaring synchrotron source with the
decreasing peak frequency. \citet{bot10shock} calculated the spectral
variations of shocks in colliding plasma shells, and reproduced the
spectral hysteresis. 

\citet{bot10shock} also simulated the optical light curves. The light
curve consists of three phases: i) a rapid rise by the onset of 
electron acceleration, ii) a gradual rise in which the shock locates
within the colliding shells, and iii) a rapid decay by synchrotron
cooling. We propose that the flares in PKS\,1749$+$096 were caused by
such internal shocks in conjunction with the variations in $\delta$,
which have a timescale analogous to that of the shock
acceleration/cooling. A similar model was proposed by
\citet{lar13s50716} for the flares of S5~0716$+$714 in which
polarization rotations were associated. In our scenario, the observed
polarization variation was dependent on the timing for the start of
acceleration/cooling relative to the time of $\theta_\mathrm{min}$. A
PA swing would only be observed when the shock acceleration starts
moderately before the source reaches the point of
$\theta_\mathrm{min}$. No flare would be observed if the cooling phase
starts much earlier before the point of $\theta_\mathrm{min}$ or if
the acceleration phase starts after it because $\delta$ is small. When
a flare starts around the point of $\theta_\mathrm{min}$, we could
observe the flare without an apparent polarization swing because
$\delta$ decreased after $\theta_\mathrm{min}$. Flares~C and D, in
which no polarization swing was detected, and Flare~G, in which $PD$
is low during the rotation event, despite having similar PAs,
may be examples of such cases. Thus, the combined effect of the
variation in $\delta$ and acceleration/cooling of electrons may be
responsible for the observed diversity of the polarization variations
in the flares. 

\subsection{Polarization features of the decay phase}

The object favors $PA$s between $-50^\circ$ and $-20^\circ$ during the
decay phase from the active states, as mentioned in section~3. This
$PA$ range is $\sim 90^\circ$ different from the $PA$ of the short
flares at maximum. In contrast to the magnetic field almost
perpendicular to the jet axis for the short flare maxima, the magnetic
field is expected to be parallel to the jet axis during the decay
phase, as illustrated in Figure~\ref{fig:hel}. The PD increased in the
early decay stage, and then decreased in the latter stage.  

In the decay phase, the electrons accelerated by the shock propagated
to the downstream region of the jet, and were significantly cooled.
The magnetic field of the source was probably not aligned by the shock
compression. Instead, the original field, which is parallel to the jet
axis, would be dominant. The electrons that were not fully cooled down
were continuously added to the downstream region throughout the active
state, which may be responsible for the observed variations in the PAs
and PDs during the decay phase. The quiescent data observed in 2014
(the blue symbols in figure~\ref{fig:rbplc}) shows a $PA$
concentration similar to the decay phase, possibly indicating a long
life-time of this component.  

\section{Summary}

Optical--NIR photo-polarimetric observations of PKS$\>$1749$+$096 in
2008--2015 were performed. We identified eight short flares having a
time-scale of a few days. The polarization features of the object are 
summarized as follows.
\begin{itemize}
\item The $PD$ tends to increase during the short flares.
\item The object favors $PA=40^\circ$--$50^\circ$ at the flare maxima,
  which is close to the position angle of the radio jet.
\item Three clear polarization rotations associated with the
  flares were detected. The other two flares also showed large $PA$
  variations during the flares.
\item The $PD$ maxima possibly precede the maxima of the total flux by
  2--4$\>$d.
\item The object favors $PA=-50^\circ$ to $-20^\circ$ in the decay
  phases from the active states.
\end{itemize}

We propose a transverse shock scenario which propagates along curved
trajectories. This scenario can explain the observed $dPA/dt$ and
time-lag between the $PD$ and flare maxima with reasonable conditions
for blazars. The shock accelerating/cooling time-scale may be
comparable to that of the change in $\delta$.

\begin{ack}
We would like to thank Gina Panopoulou for her thoughtful and detailed
comments on this paper. This work was supported by a Kakenhi
Grant-in-Aid (No. 25120007) from the Japan Society for the Promotion
of Science (JSPS). The U. of Crete group acknowledges support by the
``RoboPol project'', which is co-funded by the European Social Fund
(ESF) and Greek National Resources, and by the European Commission
Seventh Framework Programme (FP7) through grants PCIG10-GA-2011-304001
``JetPop'' and PIRSES-GA-2012-31578 ``EuroCal''.  RoboPol is a
collaboration involving the University of Crete, the Foundation of
Research and Technology-Hellas, the California Institute of
Technology, the Max-Planck Institute for Radio Astronomy, the Nicolaus
Copernicus University, and the Inter-University Center for Astronomy
and Astrophysics.
\end{ack}

\appendix
\section{TimeTubes views of the data from PKS\,1749$+$096}

Here, we introduce the TimeTubes visualization tool and an example of
its use with PKS\,1749$+$096 for future research using time-series
polarization data similar to those presented in this paper. In
TimeTubes, the trajectories of the object on the Stokes $QU$ plane are
expressed as tubes in 3D ($Q$, $U$, and time) space. The color phase
and brightness of the tubes correspond to the observed flux and color
index. The width of the tubes express the measurement errors of $Q$
and $U$. TimeTubes thus enables the behavior of six variables (flux,
color index, $Q$, $U$, and their errors) to be observed in one view. 

\begin{figure}
\begin{center}
\FigureFile(100mm,100mm){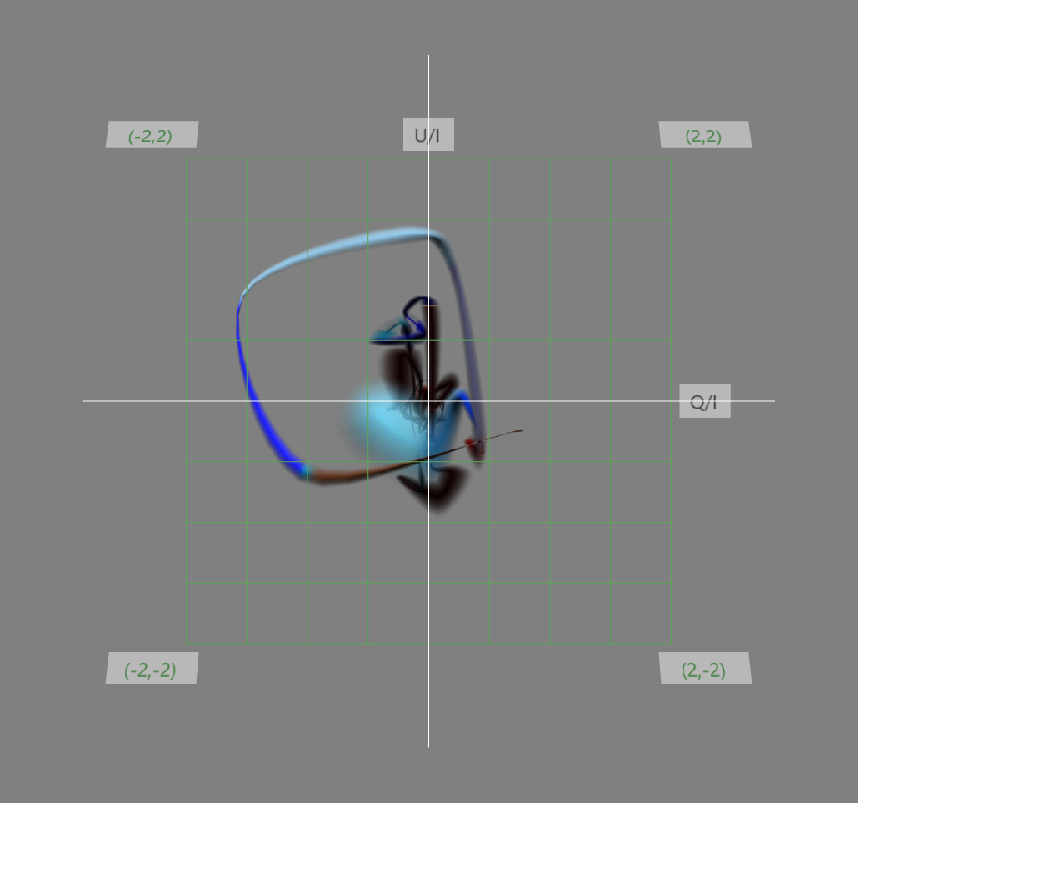}
\end{center}
\caption{TimeTubes view around flare~A.}\label{fig:tt1}
\end{figure}

\begin{figure}
\begin{center}
\FigureFile(100mm,100mm){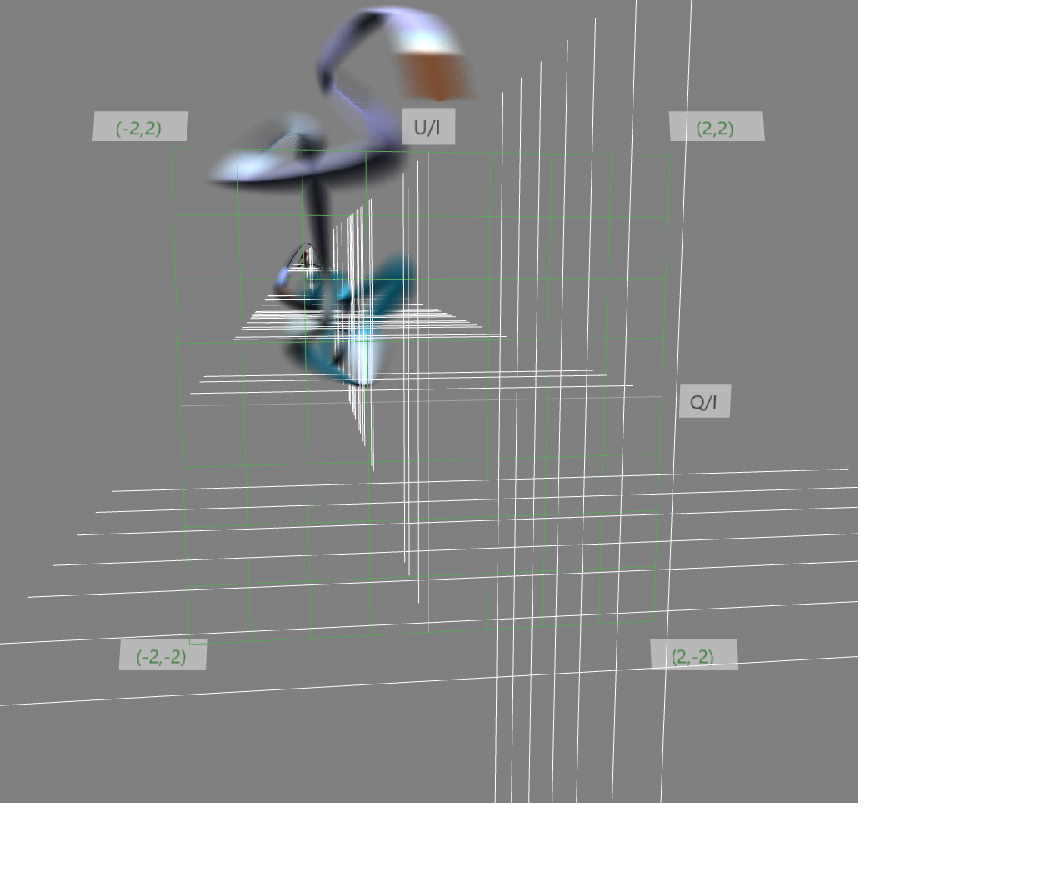}
\end{center}
\caption{TimeTubes view around flares~C and D.}\label{fig:tt2}
\end{figure}

Figure\ref{fig:tt1} shows a head-on view of TimeTubes for the data of
PKS\,1749$+$096. It shows the data around flare~A, in which a clear
polarization rotation was observed. This view indicates that the object
was first faint and red, and then became bright and blue in association
with the rotation. Figure~\ref{fig:tt2} shows a side-view of TimeTubes
around flares~C and D. Here, the object experienced two
flares, as indicated by the white color in the tube, at high $U/I$. 

TimeTubes is available at the project
site.\footnote{$\langle$http://fj.ics.keio.ac.jp/index.php/projects/spm/
  $\rangle$}



\end{document}